\documentclass[
preprint, 
 amsmath,amssymb,
 aps, physrev,
]{revtex4-2}

\usepackage{graphicx}
\usepackage{dcolumn}
\usepackage{bm}

\usepackage{graphicx} 
\usepackage{float}
\begin{document}


\title{\textbf{Preserving Hamiltonian Locality in Real-Space Coarse-Graining via Kernel Projection} 
}%

\author{Haoyuan Sun}
 \email{argv@stu.xjtu.edu.cn}
\affiliation{
 Xi'an Jiaotong University, Xi'an 710049 China
}

\date{\today}

\begin{abstract}
Numerical simulations of critical lattice systems are fundamentally limited by critical slowing down, as long-range correlations are typically established through slow temporal equilibration. A physically constrained generative framework that replaces temporal relaxation with a spatial projection mechanism for critical systems is proposed. Using the two-dimensional Ising model at criticality as a benchmark, we introduce an energy-constrained kernel that synthesizes large-scale configurations from compact equilibrated seeds by enforcing Hamiltonian-level observables. The generated configurations are projected onto the nearest-neighbor energy manifold, ensuring thermodynamic consistency while retaining universal critical properties. We show that the resulting configurations reproduce scale-invariant spin correlations, Binder cumulants, and isotropic structure factors for lattice sizes exceeding $10^6 \times 10^6$, without iterative Monte Carlo equilibration. While not a strict renormalization group transformation, and motivated by renormalization ideas, the method provides a practical inverse mapping that retains universal features of criticality and enables efficient GPU-parallel generation of ultra-large critical ensembles.

\end{abstract}

\maketitle

\section{\label{sec:level1}Introduction}

Numerical simulations of critical systems are fundamentally challenged by the emergence of long-range correlations, which are traditionally established through slow temporal equilibration. In lattice models near continuous phase transitions, this difficulty manifests as critical slowing down, where the characteristic relaxation time grows algebraically with system size. As a result, generating statistically independent equilibrium configurations at large scales remains a central bottleneck in computational statistical physics.

As a concrete benchmark, we focus on the two-dimensional nearest-neighbor Ising model at criticality, a canonical system exhibiting scale invariance and power-law spin correlations,
\begin{equation}
G(r) = \langle s_i s_{i+r} \rangle - \langle s_i \rangle \langle s_{i+r} \rangle \sim r^{-1/4}
\label{eq:correlation}
\end{equation}

Conventional Markov chain Monte Carlo (MCMC) approaches address this problem through stochastic temporal evolution, yielding dynamical exponents that severely limit scalability near criticality. Although cluster-update algorithms, such as the Wolff algorithm, can significantly reduce the dynamical exponent \cite{Wolff1989} , their inherently nonlocal updates and sequential data dependencies make them difficult to efficiently parallelize on modern accelerator hardware. This tension highlights a deeper limitation of temporal-equilibration–based approaches: long-range correlations are constructed dynamically rather than geometrically, constraining both efficiency and scalability.

The neural-network renormalization group (NNRG) framework introduced by Li and Wang \cite{LiWangNNRG} demonstrated that deep neural architectures can effectively capture long-range correlations in critical systems.Related developments have further explored how machine learning can identify relevant degrees of freedom and coarse-graining structures in many-body systems,without explicitly constructing microscopic renormalization procedures~\cite{KochJanusz2018}. Motivated by this insight, we introduce an energy-constrained generative kernel that synthesizes large-scale configurations by projecting generated fields onto the nearest-neighbor energy manifold.

While exact renormalization procedures generally generate higher-order and longer-range interactions, our approach restricts the generated configurations to the nearest-neighbor manifold by enforcing Hamiltonian-level observables. Rather than aiming to reproduce a full renormalization group flow \cite{KochJanusz2018} , the kernel provides a practical inverse mapping that preserves universal critical properties at the level of observables, enabling the scalable construction of large-scale configurations from compact equilibrated seeds.

By formulating configuration generation as a spatial projection problem, the framework naturally exploits GPU parallelism and achieves substantial performance gains over conventional Monte Carlo and cluster-update algorithms for ultra-large lattice sizes.

\begin{figure}[hb]
    \centering
    \includegraphics[width=0.8\linewidth]{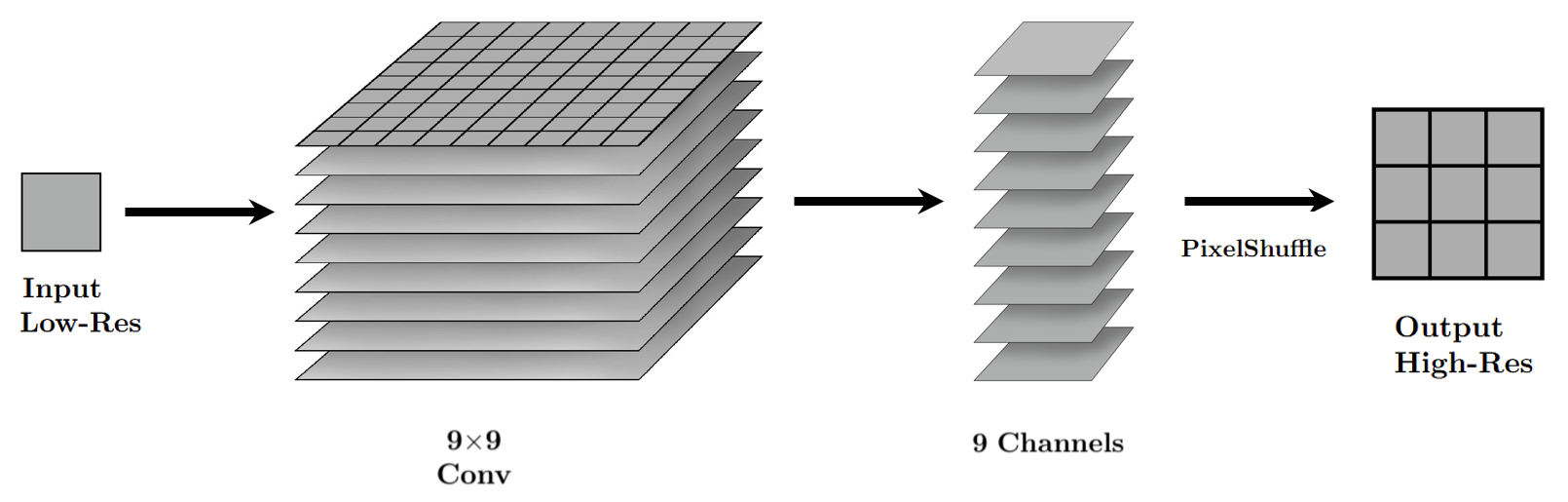}
    \caption{Schematic architecture of the ECMK.}
    \label{fig:ecmk_architecture}
\end{figure}

\section{Methodology}
\subsection{The Energy-Constrained Mapping Kernel}

To preserve physical interpretability within the coarse-graining mapping, we introduce the Energy-Constrained Mapping Kernel (ECMK). This framework centers on a projection operator $\mathcal{K}$ designed to steer the Hamiltonian's coupling projection onto the nearest-neighbor manifold.This structural constraint is enforced by optimizing learnable parameters $\{ \mathcal{W}, w_1, A \}$ such that the generated field adheres to the exact energy density at the critical temperature:

\begin{equation}
\langle \hat \epsilon \rangle = \langle \hat s_i \hat s_{i+\delta} \rangle|_{T=T_c} \approx \frac{\sqrt{2}}{2}
\end{equation}

where the hat notation signifies the estimated values produced by the neural kernel and $\delta$ implies displacements by the nearest-neighbor.

We apply a convolutional layer consisting of 9 independent $9 \times 9$ kernels. The 9-channel feature map is then processed by a PixelShuffle (PS) operator. Which maps the tensor of shape $(L, L, 9)$ to a shape of $(3L, 3L, 1)$, as shown in ~\ref{fig:ecmk_architecture}.

In practice, we simulate the coarse-graining step by applying local pooling to high-resolution Ising configurations $s_{HR}$ equilibrated at $T_c$. The ECMK then performs a high-fidelity reconstruction through a non-linear projection:

\begin{equation}
\hat{s}_{High-Res} = \text{Clamp} \left( w_1 \cdot \text{PS} (\mathcal{W} \circledast s_{Low-Res}), -A, A \right)
\end{equation}

where $\text{PS}$ denotes the PixelShuffle operator, which rearranges the $b^2$ output channels into a $b \times b$ spatial block.

\subsection{Loss Function and Physical Constraints}

To ensure that the synthesized lattice is residing in the nearest-neighbor manifold, we define a composite loss function $\mathcal{L}_{total} = \mathcal{L}_{pixel} + \gamma \mathcal{L}_{phys}$.The first term, $\mathcal{L}_{pixel}$, is the Mean Squared Error (MSE) between the predicted continuous field and the ground truth spins, ensuring local structural fidelity:

\begin{equation}
\mathcal{L}_{pixel} = \frac{1}{N} \sum_{i} (\hat{s}_{i} - s_{i, truth})^2
\end{equation}

The second and more critical term, $\mathcal{L}_{phys}$, enforces the energetic constraints of the Hamiltonian. This term is critical to make the generated Ising model "Ising". By calculating the nearest-neighbor correlation (energy density) $c_{10}$ of the generated field, we penalize deviations from the theoretical value:

\begin{equation}
\mathcal{L}_{phys} = \left( \langle \hat{s}_i \cdot \hat{s}_{i+\delta} \rangle - \epsilon_{T_c} \right)^2
\end{equation}

where $\epsilon_{T_c} \approx 0.7071$ is the target energy density. In our implementation, we deliberately set the hyperparameter $\gamma = 5000$ to balance local reconstruction and global thermodynamic constraints. 

\section{Results and Physical Verifications}

\subsection{Morphological Analysis and Fractal Visualization}

\begin{figure}
    \centering
    \includegraphics[width=0.3\linewidth]{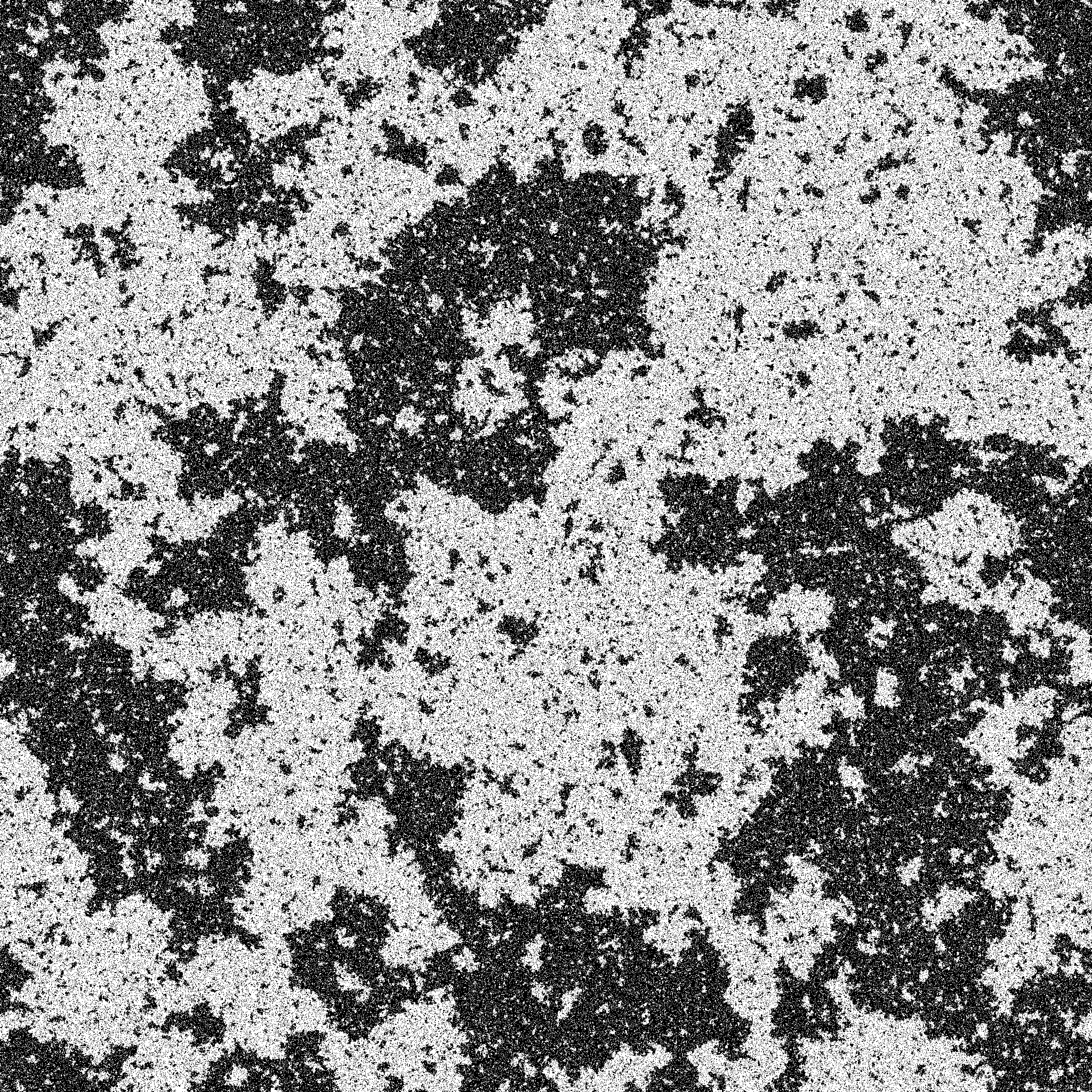}
    \caption{Representative spin configuration of the 2D Ising model at the critical temperature $\beta_c$, generated via the ECMK framework ($L=13,824$ lattice). This exhibits characteristic multi-scale domains and fractal structures of the model. One can observe the nested hierarchy of spin clusters, qualitatively demonstrating the scale-invariance property of the system near the phase transition.}
    \label{fig:ising_fig}
\end{figure}

The performance of the ECMK framework is first demonstrated through its ability to synthesize massive, physically consistent spin configurations. Starting from a $512^2$ seed to the final $13,824^2$ configuration, the model expands the lattice through three cascading stages, as shown in ~\ref{fig:ising_fig}.A brief checkerboard Monte Carlo refinement of 25 steps is applied after each expansion stage.This refinement is not intended to equilibrate the system or generate long-range correlations,but solely to suppress high-frequency local artifacts introduced by the projection operation.We have empirically verified that increasing the number of refinement steps beyond this value does not produce observable changes in long-range quantities such as correlation functions
or Binder cumulants. Accordingly, 25 steps are sufficient for the purposes of this work.

\subsection{Verification of Physical Observables}

To quantitatively evaluate the physical fidelity of the configurations generated by the ECMK, we measured key thermodynamic observables across stages. Table~\ref{tab:physics_metrics}.

\begin{table}[htbp]
\centering
\caption{Physical observables and relative errors across hierarchical upscaling stages }
\label{tab:physics_metrics}
\begin{tabular}{lccccc}
\hline
\textbf{Stage} & \textbf{Resolution} & \textbf{$\langle c_{10} \rangle$ (Energy)} & \textbf{$\langle |M| \rangle$} & \textbf{Binder $U_4$} & \textbf{Err (\%)} \\ \hline
Seed           & $512^2$             & 0.700535                                  & 0.118034                       & 0.3174                & 0.93\%            \\
Stage 1        & $1536^2$            & 0.707814                                  & 0.105912                       & 0.3048                & 0.10\%            \\
Stage 2        & $4608^2$            & 0.709802                                  & 0.094934                       & 0.3025                & 0.38\%            \\
Stage 3        & $13,824^2$          & 0.710596                                  & 0.085324                       & 0.2997                & 0.49\%            \\ \hline
\end{tabular}
\end{table}

The numerical results highlight the framework's capacity for high-precision physical synthesis. A key finding is the \textbf{stability of the critical state} during the expansion. Specifically:

\begin{itemize}
    \item \textbf{Energy Preservation:} The nearest-neighbor correlation $c_{10}$ remains highly consistent with the theoretical value. This stability is expected as $c_{10}$ is directly constrained by the physical penalty term during training, confirming that the ECMK successfully anchors the generation process onto the target Hamiltonian manifold. This helps to bypass the critical slowing down when generating spin configurations of this size.
    
    \item \textbf{Emergent Magnetization Scaling:} Another observation is the spontaneous emergence of correct global statistical behavior. Although the training process does not explicitly enforce any constraints on the total magnetization, the system naturally exhibits a monotonic decrease in $\langle |M| \rangle$ from $0.1180$ to $0.0853$ as the lattice scales up. This behavior is a non-trivial reflection of finite-size scaling theory at $T_c$, demonstrating that the ECMK captures the underlying physics of critical fluctuations beyond its explicit training objectives.
    
    \item \textbf{High-Fidelity Retention of Seed Information:} The Binder cumulant $U_4$ remains stable, hovering around $0.30$ across all three expanding stages. This stability indicates that the ECMK framework preserves the physical information and ensemble characteristics inherited from the initial seed.
\end{itemize}

\begin{figure}
    \centering
    \includegraphics[width=0.7\linewidth]{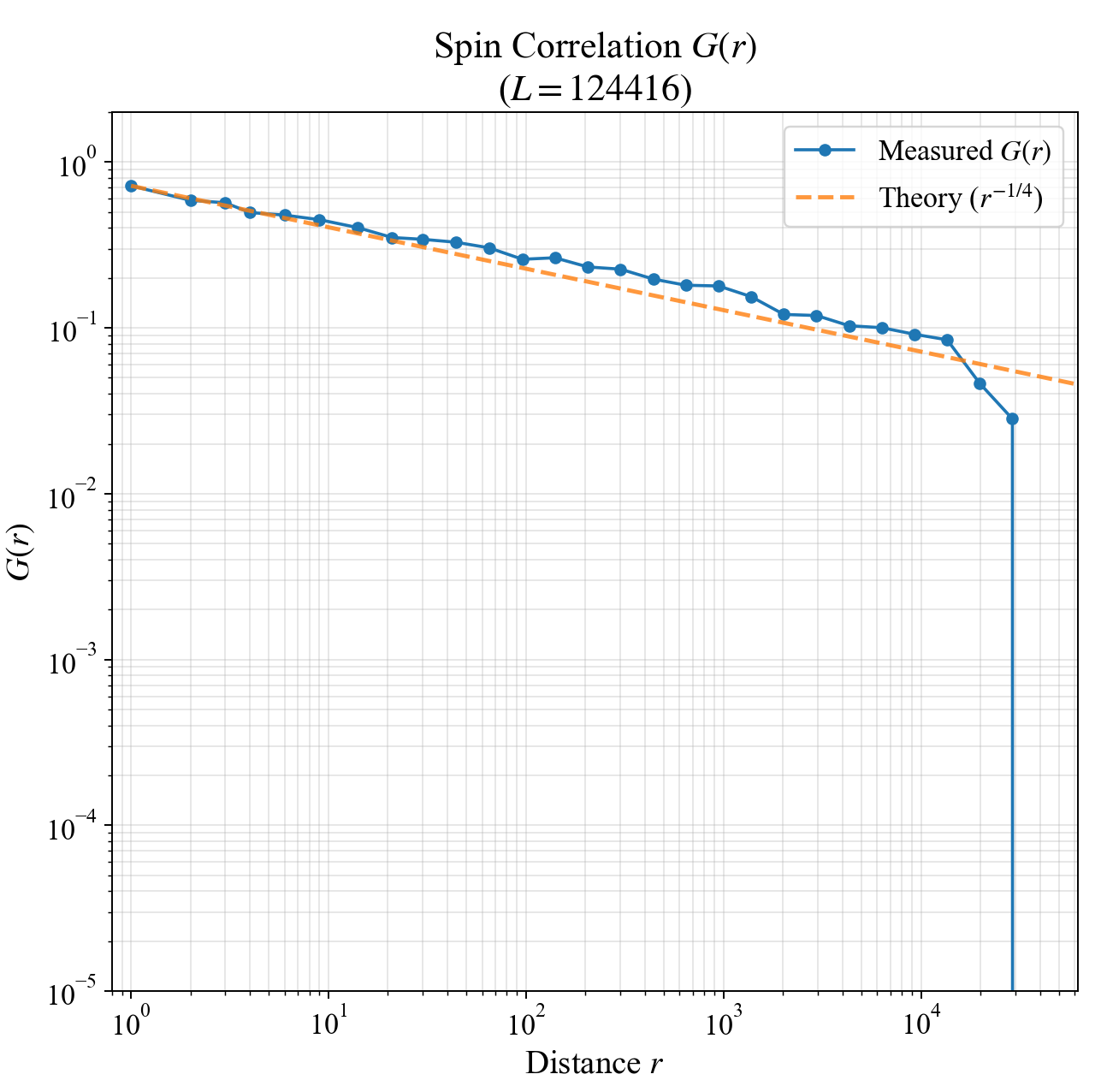}
    \caption{Spin correlation and spectral density at $L=124,416$, generated by tiling ECMK, after 4 expanding stages from the seed $L=512$}
    \label{fig:G_fig}
\end{figure}

The spatial fidelity of the ECMK is reflected in the behavior of the spin-spin correlation function $G(r)$. As illustrated in Fig.~\ref{fig:G_fig}. The measured correlation from the ultra-large $L=124,416$ lattice remains an alignment with the theoretical $r^{-1/4}$ power-law decay up to a physical distance of $r \approx 10^4$ , implying that the ECMK effectively captures and reproduces the critical correlation information at a scale way beyond the size of the kernel and the original seed itself. This indicates that the ECMK preserves and propagates critical correlations across spatial scales
far exceeding the size of the initial seed, rather than merely reproducing local correlations.

The deviation observed at $r > 10^4$ is consistent with the expected finite-size effects and the statistical resolution limit of the sampling blocks, further confirming that within the $10^0$ to $10^4$ regime, the generated configurations are consistent with the expected critical behavior of the two-dimensional Ising model at the level of measured observables.

\begin{figure}[hb]
    \centering
    \includegraphics[width=1\linewidth]{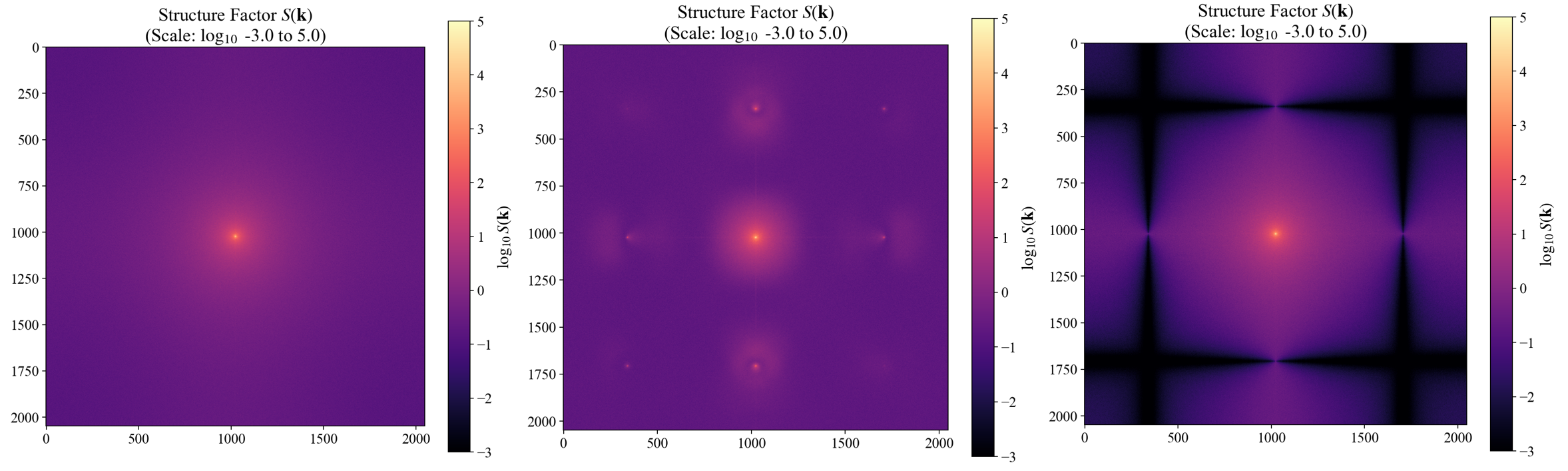}
    \caption{Comparison of static structure factors $S(\mathbf{k})$ across different synthesis methods.}
    \label{fig:sk_comparison}
\end{figure}

To further evaluate the generative quality, we perform a spectral analysis using the static structure factor $S(\mathbf{k})$, as shown in ~\ref{fig:sk_comparison}. This observable is particularly sensitive to numerical periodicities and grid-based artifacts that may be inconspicuous in real-space configurations. 

As illustrated in the comparative analysis of $S(\mathbf{k})$, the ECMK framework demonstrates significant advantages over traditional methods. Conventional expanding, such as bilinear interpolation (Fig.~\ref{fig:sk_comparison}, right), introduces severe "checkerboard" noise and spurious high-frequency peaks violating the isotropic nature of the Ising critical state. In contrast, the raw output directly from the ECMK (Fig.~\ref{fig:sk_comparison}, middle) maintains a relatively clean spectral distribution even before any stochastic refinement, implying that it is close to equilibrium directly after mapping. 

The final stage of the framework, involving a brief checkerboard Monte Carlo refinement, further polishes the fluctuations. The resulting $S(\mathbf{k})$ (Fig.~\ref{fig:sk_comparison}, left) exhibits a near-perfect isotropic scattering pattern with a sharp central peak, which is the definitive spectral signature of long-range critical fluctuations in the Ising model. The fact that the ECMK's "raw" performance already far exceeds traditional interpolation confirms its role as a high-fidelity physical field propagator.

\subsection{Computational Efficiency and Scalability}

The practical utility of the ECMK framework is most evident when comparing its computational throughput against traditional checkerboard Markov Chain Monte Carlo methods (ran on GPU) and Wolff algorithm. The speed of those algorithms are compared. For a rigorous comparison, the convergence criterion for spin configuration generations was defined by the relaxation of physical observables: the deviation from the theoretical nearest-neighbor correlation ($c_{10}$) and absolute magnetization ($|M|$) had to be less than $0.003$ and $0.01$, respectively.

\begin{table}[htbp]
\centering
\caption{Performance comparison (runtime in seconds) between traditional MCMC algorithms and the ECMK framework}
\label{tab:performance}
\begin{tabular}{lccc}
\hline\hline
Lattice Size $L$ & Cold MC (s/lattice) & Wolff (s/lattice) & ECMK (s/lattice) \\ \hline
1536  & 5.02      & 3.00      & \textbf{0.76 $\pm$ 0.04}  \\
4608  & 90.31$^a$  & 121.04    & \textbf{1.78 $\pm$ 0.14}  \\
13824 & 7296.59$^a$ & 3347.02   & \textbf{108.89 $\pm$ 26.44} \\ \hline\hline
\end{tabular}

\begin{flushleft}
\small $^a$ Did not reach physical equilibrium within the reported time. \\
\small \textit{Configuration}: Tests performed on r7-7435h CPU @ 3.1GHz and RTX 4060 GPU, representing a consumer-grade desktop PC.  ECMK results are averaged over 16 independent runs.
\end{flushleft}
\end{table}

The results, summarized in Table~\ref{tab:performance}, highlight a critical bottleneck in traditional simulations. Note that for very large scales like $L \ge 4608$, Cold MC failed to reach the defined equilibrium criteria within the recorded time, highlighting the severe impact of critical slowing down. While the Wolff algorithm effectively mitigates Critical Slowing Down, its sequential nature and complex data structures make it difficult to parallelize on GPU architectures. While as the system size increases to $L=13,824$, the ECMK achieves a speedup of approximately $31\times$ over the Wolff algorithm and over $68\times$ over the Metropolis method. By bypassing the necessity for millions of serial sweep iterations and utilizing optimized GPU tensor operations, the ECMK offers a viable path toward simulating the thermodynamic limit of statistical systems with efficiency.

The performance of ECMK stems from its expanding with a small, pre-equilibrated seed into a large-scale lattice via learned kernels, the long-range correlations are preserved and "unfolded" geometrically rather than evolved temporally. This transition from temporal evolution to spatial projection effectively sidesteps the critical slowing down, which is its fundamental departure from the local-update algorithms. The incorporation of the Hamiltonian-dependent constraint transforms the inverse coarse-graining mapping from a sequence of local operations into a globally consistent field synthesis, which is the key to the bypass of the critical slowing down.

The numerical datasets generated during this study, including the Hamiltonian matrix elements and coarse-graining kernels in \texttt{.npy} format, along with the Python implementation of the kernel projection algorithm, are openly available in Zenodo \cite{Sun2026Data}.

\section{Conclusion}
In this work, we presented the Energy-Constrained Mapping Kernel (ECMK), a physically-informed generative framework that bypasses the critical slowing down in the 2D Ising model. By shifting the paradigm from iterative temporal relaxation to spatial projection, we have demonstrated that large-scale critical configurations can be synthesized with high thermodynamic fidelity on consumer-grade hardware.

Looking forward, the ECMK framework offers a template for simulating the thermodynamic of a broader class of systems. The author will work on more non-local models to further leverage this method.

\bibliography{apssamp}

\end{document}